\documentclass[aps,prd,amsfonts,preprint, eqsecnum,nofootinbib]
{revtex4} 
\usepackage{graphics,epsfig}
\begin{document}
\preprint{WM-03-103}
%
\title{\vspace*{0.3in} Supersymmetric Flavor Models and the $B
\rightarrow \phi K_S$ Anomaly
 \vskip 0.1in}
\author{Kaustubh Agashe} \email[]{kagashe@pha.jhu.edu}
\affiliation{Department of Physics and Astronomy, Johns Hopkins
University, 3400 North Charles St., Baltimore, MD 21218-2686}
\author{Christopher D. Carone} \email[]{carone@physics.wm.edu}
\affiliation{Nuclear and Particle Theory Group, Department of Physics,
College of William and Mary, Williamsburg, VA 23187-8795}
\date{April 2003}
\begin{abstract}
We consider the flavor structure of supersymmetric theories that can
account for the deviation of the observed time-dependent CP asymmetry
in $B \rightarrow \phi K_S$ from the standard model prediction.
Assuming simple flavor symmetries and effective field theory, we
investigate possible correlations between sizable supersymmetric
contributions to $b \rightarrow s$ transitions and to flavor changing
processes that are more tightly constrained. With relatively few
assumptions, we determine the properties of minimal Yukawa and soft
mass textures that are compatible with the desired supersymmetric
flavor-changing effect and constraints. We then present explicit
models that are designed (at least approximately) to realize these
textures.  In particular, we present an Abelian model based on a
single U(1) factor and a non-trivial extra-dimensional topography that
can explain the CP asymmetry in $B \rightarrow \phi K_S$, while
suppressing other supersymmetric flavor changing effects through a
high degree of squark-quark alignment.
\end{abstract}
\maketitle

\section{Introduction}\label{sec:intro}

In the standard model (SM), the direct decay amplitudes for $B
\rightarrow J/ \psi \, K_S$ and $B \rightarrow \phi K_S$ have
approximately no weak phases in the Wolfenstein parametrization.
Thus, the CP asymmetry in both decays is due to the weak phase in $B -
\bar{B}$ mixing and hence is determined by $\sin 2 \beta$, where
$\beta$ is an angle of the unitarity triangle\footnote{In the SM, the
deviation of the CP asymmetry from $\sin 2 \beta$ in $B \rightarrow
\phi K_S$ (due to a $u$-quark penguin) is $O( \lambda^2 ) \sim 5 \%$,
where $\lambda \sim 0.22$ is the Cabibbo angle (see, for example,
reference \cite{grossman} and, for a recent study, reference
\cite{ligeti}); the deviation in the case of $B \rightarrow J / \psi
K_S$ is smaller, $O( \lambda^2 ) \times$ ratio of penguin to tree
amplitudes.}. Recently, the asymmetric $B$-factories, BABAR and BELLE,
announced that $\sin 2 \beta$ extracted from CP asymmetry in $B
\rightarrow \phi K_S$ decays (assuming the CKM paradigm for CP
violation) is smaller by $2.7 \sigma$ than the measurement using $B
\rightarrow J/ \psi \,K_S$ decays -- the world averages are $-0.39 \pm
0.41$ and $0.734 \pm 0.054$, respectively~\cite{babarbelle}. The
latter measurement of $\sin 2 \beta$ is consistent with the result of
a global fit to other data.

If this discrepancy persists, then it could be a signal of new
physics.  While the decay $B \rightarrow J / \psi \, K_S$ occurs at
tree level in the SM, the amplitude for $B \rightarrow \phi K_S$
arises at one-loop (since it involves a $b \rightarrow s s \bar{s}$
transition).  One therefore expects that new physics is more likely to
affect the latter decay mode. Also, a new physics amplitude in $B -
\bar{B}$ mixing {\em cannot} account for the anomaly since it will not
change the {\em difference} of the two CP asymmetries.  Hence, we
require a new physics contribution to the direct $B \rightarrow \phi
K_S$ decay amplitude. This new physics amplitude (a) must be
comparable to the SM amplitude and (b) have a new ${\cal O}(1)$ CP
violating phase, since the experimental CP asymmetry is $\sim 0$,
while the SM prediction is $\sim O(1)$\footnote{For recent
model-independent studies of new physics effects in this decay, see
references \cite{ciuchini, rosner}.}.

Supersymmetric extensions of the SM have new sources of flavor and CP
violation and hence are good candidates for an explanation of this
anomaly.  Consider the soft supersymmetry (SUSY) breaking masses for
the down-type squarks,
\begin{equation}
{\cal L}_{soft} \ni \tilde{d}_R^{\dagger} \left( \tilde{M}^2_d
\right)^{gauge}_{RR} \tilde{d}_R + \tilde{d}_L^{\dagger} \left( \tilde{M}^2_d
\right)^{gauge}_{LL} \tilde{d}_L + [\tilde{d}_L^{\dagger} \left( \tilde{M}^2_d
\right)^{gauge}_{LR} \tilde{d}_R + {\rm h.c.}]
\end{equation}
One may find the unitary transformations $U_{D_R}$ and $U_{D_L}$ that
diagonalize the fermion Yukawa matrices,
\begin{equation}
d^{gauge}_i = \left( U_{D} \right)_{ij} d^{mass}_j \,\,\,,
\end{equation}
and apply these transformations to the squarks so that the
quark-squark-gluino couplings are flavor-diagonal. This defines the
super-CKM basis -- a mass eigenstate basis for quarks in which the
gluino couplings are also diagonal.  The down squark mass matrices in
the super-CKM basis are given by
\begin{equation}
\left( \tilde{M}^2_d \right) ^{super-CKM}_{AB} = U_{D_A}^{\dagger}
\left( \tilde{M}^2_d \right) ^{gauge}_{AB} U_{D_B}
\label{superCKM}
\end{equation}
where the subscripts $AB$ represent either $LL$, $LR$ or $RR$.  In
general, the squark mass matrices in the super-CKM basis are {\em not}
diagonal, and provide an origin for flavor changing effects in the
low-energy theory.

\begin{figure}[t]
\epsfxsize 4 in \epsfbox{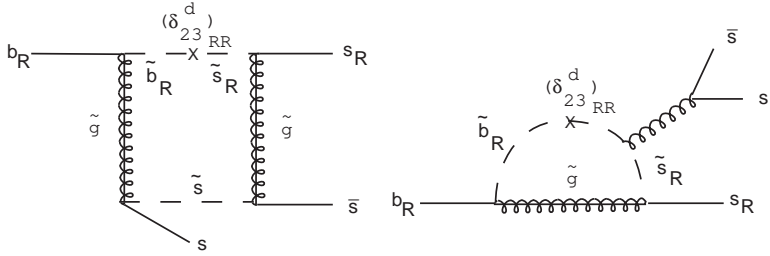} \caption{Typical direct 
supersymmetric contributions to $b\rightarrow s \bar s s$ 
transitions in the mass insertion approximation.} 
\label{fig:feynd}
\end{figure}
Flavor violation can be attributed to the ``insertion'' of an 
off-diagonal element of one of these matrices (which converts a
squark of one flavor to that of another) in a squark propagator.  If
all the squark masses are of the same order and the off-diagonal
elements in these squark mass matrices are small compared to the
diagonal elements, then this ``mass insertion'' approximation is accurate
for computing SUSY contributions to flavor changing neutral currents
(FCNC's).  In this case, convenient measures of flavor violation are
given by the parameters \cite{gabbiani}
\begin{equation}
\delta _{ij} \equiv \left( \tilde{M}^2_d \right)^{super-CKM}_{ij} /
\left( \tilde{M}^2_d \right)^{super-CKM}_{ii \; average}.
\label{delta}
\end{equation}
Possible Feynman diagrams contributing directly to $b\rightarrow 
s \bar s s$ transitions are shown in Fig.~\ref{fig:feynd}. It has 
been shown that flavor violation (in gluino-mediated Feynman 
diagrams) can explain the anomaly while satisfying the 
constraints from $b \rightarrow s \gamma$; correlation with 
$B_s-\bar{B_s}$ mixing has also been discussed~\cite{khalil, 
kane, murayama, masiero}\footnote{For an earlier study of these 
effects, see reference \cite{wyler}.  For other SUSY and non-SUSY 
explanations, see references \cite{others}.}.  Since the new 
physics amplitude has to be comparable to the SM amplitude to 
account for $B \rightarrow \phi K_S$ anomaly, it is clear that we 
need large flavor violation\footnote{The SM amplitude has a CKM 
suppression ($V_{ts}$) which roughly matches the suppression of 
the SUSY amplitude due to the fact that the superpartner masses 
are larger than the $W$, top quark masses.}, {\em e.g.}, 
$\delta^d_{23} \sim 1$ with ${\cal O}(1)$ phase\footnote{Strictly 
speaking, the mass insertion approximation is no longer valid in 
the presence of such large flavor violation. Instead, we should 
use the mass eigenstate basis for both quarks and squarks.  This 
point has been emphasized in the present context in reference 
\cite{murayama}. Since we are not concerned with a detailed 
numerical analysis in this paper, it will suffice to use the mass 
insertion approximation.}.  

In this paper, we explore {\em correlations} between a large SUSY
contribution to $B \rightarrow \phi K_S$ and the SUSY contributions to
other FCNC's.  In a model independent analysis, the various $\delta's$
can be treated as independent parameters and the stringent bounds
from, for example, $K-\bar{K}$ and $B-\bar{B}$ mixing, may be
satisfied by assumption. While this approach is important as a first
step, it is somewhat ad hoc; the nontrivial flavor structure required
of the theory strongly hints at some underlying organizing principle.
The use of flavor symmetries and effective field theory presents a 
well-motivated and consistent framework in which the question of FCNC
correlations may be addressed. Flavor symmetries explain/post-dict the
hierarchy of fermion masses and mixings while simultaneously
constraining the flavor structure of soft SUSY breaking masses.  This
in turn yields information on the pattern of SUSY contributions to
FCNC's.

While it is impossible to explore the space of possible models in 
its entirety, we can focus on the typical properties of 
successful models and let minimality guide us in finding specific 
examples. In a wide range of flavor models, one does not expect 
large contributions to $B \rightarrow \phi K_S$ from $\left( 
\delta^d_{23} \right)_{LL}$ and $\left( \delta^d_{23} 
\right)_{LR}$. One might anticipate $\left( \delta^d_{23} 
\right)_{LL} \sim \lambda^2$, where $\lambda \sim 0.22$ is the 
Cabibbo angle, since the left-handed quark mixings are of that 
order. Moreover, the $\left( \tilde{M}^2_d \right)^{gauge}_{LR}$ 
terms and the Yukawa couplings transform identically under flavor 
symmetries and are often ``aligned'', {\em i.e.}, are 
diagonalized by the same unitary transformations, so that the 
$\delta _{LR}$'s are suppressed. We therefore focus on the 
possibility that the $B \rightarrow \phi K_S$ anomaly is 
explained by a large imaginary part of $\left( \delta^d_{23} 
\right)_{RR}$, the most likely solution if $\tan\beta$, the ratio 
of Higgs vacuum expectation values (vevs), is order one.  (We do 
not consider the case of large $\tan\beta$ in this paper.)
It is worth noting that the preferred regions of supersymmetry parameter 
space that follow from the analyses in Refs.~\cite{khalil, 
kane, murayama, masiero} depend crucially on the renormalization of
the Wilson coefficients in the $\Delta B =1$ effective Hamiltonian.
The results are obtained numerically, cover only parts of the 
multidimensional supersymmetry parameter space, and cannot be summarized usefully in 
any simple analytic approximation.  We therefore use the numerical results in Figs.~3 
through 7 of Ref.~\cite{murayama} as a basis for our investigation
of more explicit supersymmetric models. In 
Section II, we study the properties of minimal mass matrix 
textures that lead to $\left(\delta^d_{23} \right)_{RR} \sim 1$, 
and determine the correlation with other $\delta$ 
parameters. In Section III, we present explicit flavor models that 
approximate these textures.

It is worth pointing out that our analysis is of interest even if the
current anomaly in $B \rightarrow \phi K_S$ turns out to be a
statistical fluctuation.  One possibility is that the SUSY
contribution in all $b \rightarrow s$ transitions is large, but
CP-{\em conserving} so that the CP asymmetry in $B \rightarrow \phi
K_S$ is not affected.  In this case, anticipated improvements in the
measurement of $B_s-\bar{B}_s$ mixing have the potential of revealing
the SUSY contribution.  Another possibility is that the SUSY
contribution is CP-{\em violating}, but is {\em small} in the
amplitude for the decay $B \rightarrow \phi K_S$ and {\em large} in
the amplitude for $b \rightarrow s \gamma$ or in $B_s-\bar{B}_s$
mixing.  This will result in a direct CP asymmetry in $B \rightarrow
X_s \gamma$ or a mixing-induced, time-dependent CP asymmetry in $B_s
\rightarrow J / \psi \, \phi$, which potentially can be detected in
ongoing or future experiments.  Whether such effects are consistent
with the tight bounds on $s \rightarrow d$ and $b \rightarrow d$
transitions in realistic models is also addressed in our approach.

\section{Texture Analysis}

In order to obtain a large value for $\left( \delta^d_{23}
\right)_{RR}$, one must either have (a) squark nondegeneracy and a
large 2-3 rotation on right-handed (RH) fields from down-quark Yukawa
diagonalization, or (b) large off-diagonal terms in the squark
mass matrix in the gauge basis. We consider these possibilities in turn.
Of course, combinations of (a) and (b) are also possible, as we will see
in Section III.

\subsection{Large RH $2-3$ rotation in down quark Yukawa matrix}

For simplicity, we assume that in the gauge basis the off-diagonal
terms in the RH down squark mass matrix are small,
\begin{equation}
\left( \tilde{M}^2_d \right) ^{gauge}_{RR} \approx \left(
\begin{array}{ccc} m^2_{\tilde{d}_R} & 0 & 0 \\ 0 & m^2_{\tilde{s}_R}
& 0 \\ 0 & 0 & m^2_{\tilde{b}_R}
\end{array} \right) \,\,\, ,
\label{squarkgauge}
\end{equation}
so that $\left( \delta^d_{23} \right)_{RR} \sim 1$ is obtained only
from large rotation between strange and bottom quarks.  Such a large
mixing might be motivated in grand unified theories as
follows~\cite{neutrino}: it is possible that the large $\nu_{\mu} -
\nu_{\tau}$ mixing required to explain the atmospheric neutrino
anomaly originates in the left-handed (LH) charged lepton mass matrix.
Since right-handed down-type quarks and left-handed leptons are part
of the same multiplet, large $b_R-s_R$ mixing is an immediate
consequence.

The ``minimal'' quark texture which gives the desired phenomenology
can be obtained as follows. We require ${\cal O}(1)$ rotation in the 2-3
sector for RH down-type quarks (with CP violation) to explain the $B
\rightarrow \phi K_S$ anomaly. Also, the size of bottom and strange
quark Yukawa couplings renormalized at a high scale are 
${\cal O}( \lambda^2 )$ and ${\cal O}(\lambda ^4 )$, respectively. Thus, 
the ``minimal'' quark texture required in $2-3$ block of the down quark
Yukawa matrix  is:
\begin{equation}
\left( Y_d \right)_{2-3 \; block} \sim \left(
\begin{array}{cc}
\lambda ^4 & \lambda ^4 \\ \lambda^2 \exp ( i \phi ) & \lambda^2
\end{array} \right).
\label{texture23}
\end{equation}
Here, we use the notation ${\cal L} \ni Y_{d \; ij} \bar{d}_{L \; i}
d_{R \; j}$, where $d_L$ and $d_R$ denote the left-handed and
right-handed down-type quarks ($i, j$ are generation indices).  Also,
unless stated otherwise, there are unknown, ${\cal O}(1)$ fluctuations in the
coefficients of all elements of Yukawa/mass matrices.  Neglecting the
first generation entries, which are typically small, the phase in
$(3,2)$ element can be ``transferred'' to the $(2,3)$ element by field
redefinitions so that the $(2,3)$ element has to be $O( \lambda^4)$ to
get the maximum CP violation. The $(2,3)$ element cannot be larger, or
it would give $V_{cb} > \lambda^2$.

In order to study the implications of the large
$2-3$ RH rotation in $Y_d$ for other FCNC's, we need the texture in the
$1-2$ block of $Y_d$, which depends on whether the Cabibbo angle originates
in $Y_d$ or in $Y_u$. We analyze these cases in turn.

\begin{center}
{\bf Cabibbo angle from $Y_d$}
\end{center}

In this case, we have $\left( U_{D_L} \right)_{12, 21} \sim \lambda$
from the $1-2$ rotation among the LH down-type quarks.  If
$m^2_{\tilde{d}_L} \approx \! \! \! \! \!  \not \; \;
m^2_{\tilde{s}_L}$, then it follows from Eqs. (\ref{superCKM}) and
(\ref{delta}) that $\left( \delta^d_{12} \right)_{LL} \sim
\lambda$. This is ruled out by the $K-\bar{K}$ mixing bound $\left(
\delta^d_{12} \right)_{LL} \stackrel{<}{\sim} 0.1-0.04$, for squark
masses $\alt 1$~TeV~\cite{gabbiani}.  Hence, to evade this bound, we
need approximate degeneracy between first and second generation
squarks. Models based on non-Abelian flavor symmetries with a ${\bf
2}+{\bf 1}$ representation structure for fields of the three
generations have this feature.  The simplest models of this type (for
example, based on U(2) symmetry~\cite{u2_1,u2o} or one of its discrete
subgroups~\cite{u2_2}) also lead to the decomposition ${\bf 2}\otimes
{\bf 2} = {\bf 3} \oplus {\bf 1}$, where the ${\bf 3}$ and ${\bf 1}$
correspond to two-by-two symmetric and antisymmetric matrices.  As a
result, the symmetry breaking entries in the 1-2 block of $Y_d$ tend
to appear in a symmetric or antisymmetric pattern.  Motivated by the
minimality of these models and their success in maintaining the squark
degeneracy that we require given our present assumptions, we combine a
U(2)-like texture in 1-2 sector with Eq. (\ref{texture23}):
\begin{equation}
Y_d \sim \left(
\begin{array}{ccc}
0 & \lambda^5 & 0 \\ \lambda^5 & \lambda^4 & \lambda^4 \\ 0 &
\lambda^2 \exp ( i \phi ) & \lambda^2
\end{array} \right)
\label{eq:fullyd}
\end{equation}
Generally, the $(1,2)$ and $(2,1)$ entries have contributions from
symmetry breaking fields, flavons, that transform as symmetric or
antisymmetric tensors under the flavor group.  Thus, unless symmetric
and antisymmetric flavon vevs are fine tuned, the $(1,2)$ and $(2,1)$
entries will be of the same size, namely $O(\lambda ^5)$.  These
entries provide an origin for the Cabibbo angle, and also yield a
down quark Yukawa coupling of the right order, namely $\lambda ^6$.

To proceed further in correlating the large $2-3$ rotation with other
FCNC's, we obtain the form of $U_{D_R}$.  In general, it is difficult
to analytically diagonalize $3 \times 3$ matrices. However, if the
entries in the Yukawa matrices have a hierarchical structure, then the
matrices can be diagonalized perturbatively by three successive
rotations in the 2-3, 1-3 and 1-2 subspaces (see, for example,
reference \cite{rasin}).  Denoting the sines of the rotation angles by
$s^{R (L)}_{23}$, $s^{R (L)}_{13}$ and $s^{R (L)}_{12}$ for the RH
(LH) quarks, we have
\begin{equation}
Y_d^{diagonal} = \left( U_{D_L} \right)^{\dagger} Y_D U_{D_R},
\end{equation}
where
\begin{equation}
U_{D_R} \approx \left( \begin{array}{ccc} 1 & 0 & 0 \\ 0 & 1 & s^R_{23} \\ 0
& - s^R_{23} & 1
\end{array}
\right) \left( \begin{array}{ccc} 1 & 0 & s^R_{13}\\ 0 & 1 & 0 \\ -
s^R_{13} & 0 & 1
\end{array}
\right) \left( \begin{array}{ccc} 1 & s^R_{12} & 0 \\ - s^R_{12} & 1 &
0 \\ 0 & 0 & 1
\end{array}
\right) \,\,\, .
\label{DR}
\end{equation}
A similar decomposition is valid for $U_{D_L}$.  Given the texture in
Eq.~(\ref{eq:fullyd}), it is not hard to see that $s^R_{12} \sim {\cal
O}(\lambda)$, $s^R_{23} \sim {\cal O}(1)$ and that $s^R_{13}$ is very
small\footnote{Strictly speaking, although perturbative diagonalization
cannot be used in this case since $2-3$ mixing is ${\cal O}(1)$, it suffices
for our semi-quantitative analysis.}. Expanding the RHS of
Eq. (\ref{DR}), we get $\left( U_{D_R} \right)_{12} \sim s^R_{12}$,
$\left( U_{D_R} \right)_{21} \sim s^R_{12} + s^R_{13} s^R_{23}$,
$\left( U_{D_R} \right)_{23} \sim s^R_{23}$ and $\left( U_{D_R}
\right)_{32} \sim s^R_{23} + s_{13}^R s^R_{12}$ so that $\left(
U_{D_R} \right)_{12, 21} \sim \lambda$ and $\left( U_{D_R}
\right)_{23, 32} \sim O(1)$, as one might expect.  What is more
interesting is the observation that
\begin{equation}
\left(U_{D_R} \right)_{31} = -s^R_{13}+s^R_{23} s^R_{12} \,\,\, ,
\label{induced13}
\end{equation}
which is enhanced by the large 2-3 mixing.  Using the above values of
the $3$ rotations, we see that perturbative diagonalization gives
$\left( U_{D_R} \right)_{31} \sim \lambda$.  We have confirmed this
expectation by a numerical analysis, which also shows that $\left(
U_{D_R} \right)_{31}$ has a phase\footnote{We work in the basis where
the quark masses are real and the CKM matrix, i.e., $U^{\dagger}_{U_L}
U_{D_L}$ has the standard form given in the Review of Particle
Physics~\cite{rpp}.}.  Thus, we obtain the following significant
result: $\left(U_{D_R} \right)_{31} \sim {\cal O}(\lambda)$ is a
generic consequence of the minimal texture in Eq.~(\ref{eq:fullyd}),
even though the 1-3 rotation defined by Eq.~(\ref{DR}) is small.

Let  us now estimate the relevant $\delta$ parameters.  In non-Abelian 
models with a ${\bf 2} \oplus {\bf 1}$ representation structure, one 
expects that $m_{\tilde{d}_R} \approx m_{\tilde{s}_R}$ while 
$m_{\tilde{d}_R}$ and $m_{\tilde{b}_R}$ are of the same order but not 
degenerate. Then, using $\left( U_{D_R} \right)_{32, 33} \sim 1$ and 
$\left(U_{D_R} \right)_{31} \sim \lambda$ in Eq. (\ref{superCKM}), we find
$\left(\delta ^d _{12} \right)_{RR} \sim \lambda$ (with no phase) and
$\left( \delta ^d _{13} \right)_{RR} \sim \lambda \exp ( i \phi)$.  On
the other hand, the limits from $K - \bar{K}$ and $B - \bar{B}$ mixing
require $\hbox{Re} \left( \delta ^d _{12} \right)_{RR}
\alt 0.04$ and $\hbox{Re} \left( \delta ^d _{13}
\right)_{RR} \alt 0.1$, respectively, for squark and
gluino masses $\approx 500$ GeV~\cite{gabbiani}.  Thus, the SUSY
contribution to $\Delta m_K$ is too large with this texture, but there
is no SUSY contribution to $\epsilon_K$. The SUSY contribution to
$\Delta m _B$ is borderline ({\em i.e.}, comparable to the
experimental value) and has an ${\cal O}(1)$ phase. Thus, the
time-dependent CP asymmetry in $B
\rightarrow J / \psi K_S$ measures a combination of angle $\beta$ and
the new phase $\phi$.  

Clearly, our theory must be modified if the contribution to $K - \bar{K}$ 
mixing is to be sufficiently suppressed. One could entertain the possibility 
that the squarks of the first two generations are much heavier than the 
third;  a squark mass spectrum of the form $m_{\tilde{d},\tilde{s}_R} 
\agt 1$~TeV and $m_{\tilde{b}_R} \sim$ a few 
$100$~GeV can still explain the $B \rightarrow \phi K_S$ 
anomaly~\cite{murayama}. To study this case, one must work in the mass
eigenstate basis for both the quarks and squarks.  Given the assumed
form of Eq.~(\ref{squarkgauge}), the gauge and mass bases for the squarks
are the same; the quark-squark-gluino coupling in the mass basis is
then given by
\begin{equation}
\tilde{g} \left( \tilde{d}^{mass}_i \right)^{\dagger} \left( U_{D_R}
\right)_{ij} d^{mass}_j  \,\,\, .
\label{coupling}
\end{equation}
Dominant contributions to FCNC's come from Feynman diagrams involving
the exchange of the light $\tilde{b}_R$. From Eq. (\ref{coupling}), we
see that its couplings to $d$ and $s$ quarks are given by $\left(
U_{D_R} \right)_{31} \sim \lambda$ (with a phase) and $\left( U_{D_R}
\right)_{32} \sim 1$ (with a phase), respectively. Box diagrams
involving the exchange of $\tilde{b}_R$ and a gluino generate a
CP-conserving, $\Delta S = 2$ $4$-fermion operator with coefficient
$\propto \lambda ^2 / m^2_{\tilde{b}_R}$, and a $\Delta B = 2$
$4$-fermion operator with a coefficient of the same order, but with a
phase.  It is easy to see that this is equivalent to having $\left(
\delta^d_{12,13} \right)_{RR} \sim \lambda$ and thus gives too large a
contribution to $\Delta m _K$ and a large CP-violating contribution to
$B - \bar{B}$ mixing, just as before.

If one, on the other hand, tries to make {\em all} the quarks heavier
than a TeV, than one must take the gluino mass to be relatively light,
(a few $100$ GeV) to affect $B \rightarrow \phi K_S$ (see
Ref.~\cite{murayama}).  Thus, $x \equiv m^2_{\tilde{g}} /
m^2_{\tilde{q}} \ll 1$. Heavy squarks relax the bounds on $\delta$'s,
whereas small $x$ tightens the bounds. As a result, the limits on
$\delta$'s stay about the same \cite{gabbiani}. Thus, the large
contributions to $K - \bar{K}$ and $B - \bar{B}$ are not alleviated if
we are also to explain the anomaly of interest.

These considerations lead us to the conclusion that we must suppress
the $(2,1)$ entry of Eq.~(\ref{eq:fullyd}). For U(2)-like textures, with
symmetric and antisymmetric contributions to the upper two-by-two block of
$Y_d$, this suggests a mild fine tuning is required.  We will construct
an explicit model based on this solution in Section III.

\begin{center}  
{\bf Cabibbo angle from $Y_u$}
\end{center}

For models in which the Cabibbo angle originates in $Y_u$, we can 
we can choose a minimal block diagonal form for $Y_d$:
\begin{equation}
Y_d \sim \left(
\begin{array}{ccc} 
\lambda^6 & 0 & 0 \\ 0 & \lambda^4 & \lambda^4 \\ 0 & \lambda^2 \exp
(i \phi) & \lambda^2
\end{array} \right) \,\,\, .
\label{Cabibboup1}
\end{equation}
To generate the Cabibbo angle, we must introduce a $(1,2)$ entry of the
appropriate size in $Y_u$:
\begin{equation}
Y_u \sim \left(
\begin{array}{ccc} 
\lambda^8 & \lambda^5 & 0 \\ 0 & \lambda^4 & \lambda^2 \\ 0 & 0 & 1
\end{array} \right) \,\,\, .
\label{Cabibboup2}
\end{equation}
Note that the eigenvalues of $Y_u$ are in the correct ratio for the up-type
quarks, namely $1$::$\lambda^4$::$\lambda^8$.
If we were to follow the same non-Abelian ansatz described earlier, we 
would also expect an ${\cal O}(\lambda^5)$ $(2,1)$ entry in $Y_u$.
However, this would lead to an up quark Yukawa coupling of order 
$\lambda^6$, about a factor of $25$ too large.  The fine tuning required
to fix this problem is much greater than what we needed to adequately
suppress the flavor changing problems in our previous example.  So it is
more natural in this case to assume that the $(2,1)$ entry in $Y_u$ is 
absent, and to discard our assumption that the underlying theory is 
non-Abelian\footnote{Of course, the underlying symmetry could still be
non-Abelian but of a more complicated form than we have assumed, {\em e.g.},
products of simple non-Abelian factors. We do not consider this possibility
here.}; unwanted SUSY FCNC's must then be suppressed
by a suitable quark-squark alignment.  We will show in Section III that
this can be achieved in Abelian models with multiple U(1) factors
(a la Nir-Seiberg~\cite{NS}), or in a new class of models involving 
fewer U(1)'s and a nontrivial extra-dimensional topography. 

From the texture of $Y_d$ and the form of the squark mass matrix in 
Eq. (\ref{squarkgauge}), we see that the first generation is decoupled 
from the second and third in the down sector.  Thus, there are 
no SUSY contributions to $K-\bar{K}$ and $B-\bar{B}$ mixing, even if the 
three down-type squarks are not degenerate. However, since
$\left( U_{U_L} \right)_{12} \sim \lambda$,  some mild degeneracy
between $m_{\tilde{u}_L}$ and $m_{\tilde{c}_L}$ may be required to satisfy
the bounds on $D-\bar{D}$ mixing.

\subsection{Large 2-3 mixing in RH down squark mass matrix}

In this case, the ``minimal'' texture for down-type squark mass matrix in 
the gauge basis is
\begin{equation}
\left( \tilde{M}^2_d \right) ^{gauge}_{RR}
\propto
\left(
\begin{array}{ccc}
1 & 0 & 0 \\
0 & 1 & \exp (i \phi) \\ 0 & \exp (-i \phi) & 1
\end{array} 
\right),
\label{squarkmixing}
\end{equation}
where, for simplicity, we have assumed that the off-diagonal elements
involving down squarks are small. In the discussion that follows
immediately below, we assume that large 2-3 mixing is present only in
the supersymmetry breaking sector of the theory. However, it should be
pointed out that in most realistic models, the form of
Eq.~(\ref{squarkmixing}) suggests that the entire 2-3 block is flavor
group invariant, and that $s_R$ and $b_R$ have the same flavor
charges. In this case, one expects large 2-3 mixing from both the
Yukawa and soft mass matrices, as illustrated in the first model of
Section III. 

If the Cabibbo angle comes from the down sector, then as before we
need degeneracy of $m_{\tilde{d}_L}$ and $m_{\tilde{s}_L}$ and hence a
non-Abelian flavor symmetry.  Given our previous assumptions on the 
symmetry or antisymmetry of the upper two-by-two block of $Y_d$, we
expect generically that  $\left( U_{D_R} \right)_{12, 21} \sim
\lambda$.  However, unlike our earlier example, 
$\left( U_{D_R} \right)_{31}$ can be very small (see Eq. (\ref{induced13})) 
since we no longer need an ${\cal O}(1)$ 2-3 RH rotation in the quark 
sector. Using Eq.~(\ref{superCKM}) and $\left( U_{D_R} \right)_{12, 21} \sim
\lambda$, we get
\begin{equation}
\big[\left( \tilde{M}^2_d \right)
^{super-CKM}_{RR}\big]_{13} \sim \lambda \times 
\big[\left( \tilde{M}^2_d \right) ^{gauge}_{RR} \big]_{23}
\end{equation}
resulting in $\left( \delta ^d_{13} \right)_{RR} \sim \lambda$ 
(with a phase), and large $B - \bar{B}$ mixing as before. 
Since 
$\left( U_{D_R} \right)_{31}$ and $\left(U_{D_R} \right)_{32}$ can be
small
and $m_{\tilde{d}_R}
\approx m_{\tilde{s}_R}$ is implied by the non-Abelian flavor
symmetry, we can check (using Eq. (\ref{superCKM}))
that $\left(
\delta^d_{12} \right)_{RR}$ and hence $K - \bar{K}$ mixing can be
small, unlike our earlier example.

If the Cabibbo angle originates in the up-quark sector, then the
off-diagonal entries of $Y_d$ involving the down quark can be small
or zero.  Assuming further that the down squark mass matrix is as in
Eq. (\ref{squarkmixing}), we see that the down quark is decoupled from
the bottom and strange quarks so that there are no SUSY contributions
to $K-\bar{K}$ and $B-\bar{B}$ mixing.

\section{Models}
In this section, we will consider simple Abelian and non-Abelian
models that can account for the discrepancy in the measured value of
$\sin 2\beta$.  Our Abelian model will realize the Yukawa textures in
Eqs. (\ref{Cabibboup1}) and (\ref{Cabibboup2}) by construction, just
as the well-known alignment models of Ref.~\cite{NS}. However, unlike
these models, we rely on localization in extra dimensions rather that
holomorphy to achieve an appropriate quark-squark alignment for the
lighter generations.  This allows us to achieve a viable phenomenology
using only one flavor U(1) factor.  An interesting feature of our
model is that texture zeros can occur in non-holomorphic terms, such
as the soft scalar masses, as a consequence of extra-dimensional
locality.  This leads to somewhat different scalar mass matrix
textures -- exact zero entries for the off-diagonal elements involving
the down-squark, where they otherwise would be finite but small.  In
our model and those of Ref.~\cite{NS},
$\left(\delta^d_{23}\right)_{RR}$ is due to a large RH 2-3 rotation in
the down quark Yukawa matrix {\em and} large 2-3 mixing in the down
squark mass matrix. In the non-Abelian example, we will show how a
U(2) flavor model may be modified to provide for the desired $b$-$s$
mixing; large $K$-$\bar K$ mixing can be avoided provided that a mild
fine-tuning is allowed.

\subsection{An Extra-dimensional Abelian Model}

In this model, we assume the horizontal flavor symmetry $G_f=$~U(1);
the particle content is that of the minimal supersymmetric standard
model with one additional field $\phi$ that has U(1) charge $+1$.  We
assume that the ratio of the vev of $\phi$ to the ultraviolet cutoff
of the effective theory, $M_f$, is approximately given by the Cabibbo
angle,
\begin{equation}
\frac{\langle \phi \rangle}{M_f} = \lambda \approx 0.22 \,\,\, .
\label{eq:spurion}
\end{equation}
We assume that $M_f$ is generically well below the fundamental
gravitational scale $M_*$ and that $M_f$ sets the mass scale for all
the operators contributing to the flavor structure of the theory. The
U(1) charge assignments of the matter fields are given as follows:
\begin{equation}
\begin{array} {lcl} E \sim Q \sim (-3, -2, \, 0), & \,
\,\,\,\,\,\, & L \sim D \sim (-3 -2 -2), \\ U \sim (-5, -2,\, 0), & &
H_U, H_D \sim 0\,. \end{array}
\end{equation}
Yukawa textures originate as higher-dimensional operators involving
the MSSM matter fields and powers of $\langle \phi \rangle/M_f$.  The
model described thus far is not viable since the Cabbibo angle
receives an unsuppressed contribution from the diagonalization of the
down quark Yukawa matrix (as one can verify from the explicit form for $Y_d$), 
leading to $(\delta_{12}^d)_{LL}\sim \lambda$; this exceeds the bounds 
from $K^0$-$\bar K^0$ mixing and $\epsilon_K$ if superparticle masses 
are a few hundred GeV.

\begin{figure}[ht]
\epsfxsize 3.3 in \epsfbox{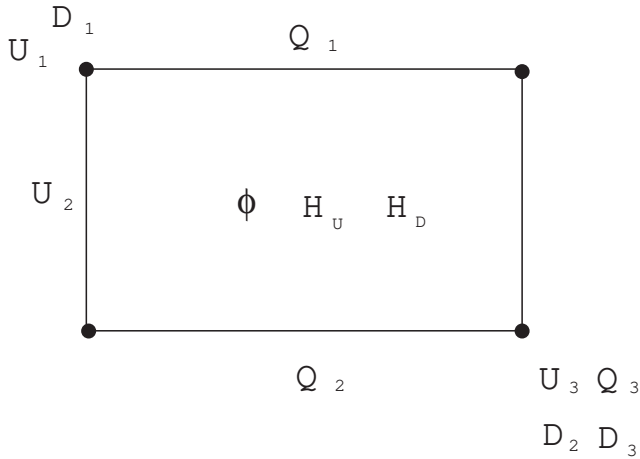}
\caption{Extra-dimensional topography for the Abelian model.  Corners
represent fixed points, sides represent fixed lines, and the interior
represents the entire bulk.}
\label{fig:rect}
\end{figure}

We now consider a possible non-trivial, extra-dimensional topography
for the model.  We assume that there are two extra spatial dimensions,
compactified on the orbifold $(S^1/Z_2)^2$.  We take the
compactification radii to be the same, namely $R$, so that fixed
points exist at $(0,0)$, $(0,\pi R)$, $(\pi R, 0)$ and $(\pi R, \pi
R)$.  The space can be described as a rectangle, with fields confined
either to corners (fixed points of both $S^1/Z_2$ factors), sides
(fixed point of one $S^1/Z_2$ factor), or defined everywhere (the
six-dimensional bulk).  We choose to localize our fields as shown in
Fig.~\ref{fig:rect}.  In addition, we assume that the unknown, high-energy
dynamics that is responsible for generating $M_f$-suppressed operators is
localized at fixed points of both $S^1/Z_2$ factors.

Given this construction, the separation of fields at isolated fixed
points prevents Yukawa couplings involving these fields in the
four-dimensional theory, after dimensional reduction.  For the
couplings that do remain, one must generally take into account volume
suppression factors ({\em i.e.}, powers of the ratio $R^{-1}/M_f$)
that, to varying degrees, suppress the effective four-dimensional
interactions. In our model we will assume that $R^{-1} \sim M_f$ so
that these factors are ${\cal O}(1)$ and do not alter the analysis.
Moreover, we will take $R^{-1}$ to be sufficiently large so that only
the physics of the zero-modes of the MSSM matter fields will be
relevant to the low-energy phenomenology.  Finally, a minor technical
point: for MSSM matter fields that are confined to 5D subspaces of the
6D bulk, we allow for the introduction of chiral-conjugate mirror
fields with opposite parity under the appropriate $Z_2$ factor so that
chiral zero modes can be obtained.
 
Given the constraints of symmetry and geometry, we find the following
textures for the Yukawa matrices:
\begin{equation}
Y_u \sim \left( \begin{array}{ccc} \lambda^8 & \lambda^5 & 0 \\ 0 &
\lambda^4 & \lambda^2 \\ 0 & 0 & 1 \end{array}\right),
\,\,\,\,\,\,\,\,\, Y_d \sim \left( \begin{array}{ccc} \lambda^6 & 0 &
0 \\ 0 & \lambda^4 & \lambda^4 \\ 0 & \lambda^2 & \lambda^2
\end{array}\right) \,\,\, .
\label{eq:abyuk}
\end{equation}
While these are the same as in Ref.~\cite{NS} (by construction), the
soft mass squared matrices are somewhat different:
\begin{equation}
m^2_Q \sim m^2 \left( \begin{array}{ccc} 1 & 0 & 0 \\ 0 & 1 &
\lambda^2 \\ 0 & \lambda^2 & 1 \end{array}\right), \,\,\,\,\,\, m^2_D
\sim m^2 \left( \begin{array}{ccc} 1 & 0 & 0 \\ 0 & 1 & 1 \\ 0 & 1 & 1
\end{array}\right), \,\,\,\,\,\, m^2_U \sim m^2 \left(
\begin{array}{ccc} 1 & \lambda^3 & 0 \\ \lambda^3 & 1 & 0 \\ 0 & 0 & 1
\end{array}\right) \,\,\, .
\label{softtexture}
\end{equation}
As far as strangeness changing processes are concerned, this model
yields a remarkable level of quark-squark alignment; notice that there
are no 1-2 down quark rotations (on either left- or right-handed
fields), and no 1-2 entries in either $m^2_Q$ or $m^2_D$!  The model
provides the desired ${\cal O}(1)$ 2-3 right-handed down-quark mixing
with an irremovable phase (as well as large 2-3 mixing in $m^2_D$),
and like other models of this type predicts $D^0$-$\bar D^0$ mixing
near the experimental limit.  It is worth noting that a choice of
$M_f$ closer to $M_*$ may lead to exponentially suppressed, though
non-negligible, interactions between fields at isolated fixed points,
due to the exchange of string states.  Such effects are not relevant
here given our choice of scales, but could be of interest in similar
flavor models.

\subsection{A Non-Abelian Example}

Non-Abelian flavor symmetries resolve the supersymmetric flavor
problem by maintaining a sufficient degree of degeneracy among squarks
of the first two generations.  Theories of flavor based on U(2)
symmetry assume a ${\bf 2}+{\bf 1}$ representation structure for the
three generations, and a two-stage symmetry breaking
\begin{equation}
U(2) \stackrel{\epsilon}{\rightarrow} U(1)
\stackrel{\epsilon'}{\rightarrow} \mbox{ nothing} \,\, ,
\label{eq:brk}
\end{equation}
where the U(1) factor represents phase rotations on fields of the
first generation, and where $\epsilon$ and $\epsilon'$ are small
parameters defined in analogy to Eq.~(\ref{eq:spurion}).  For a
detailed discussion of models based on U(2) symmetry and its discrete
subgroups we refer the reader to Ref.~\cite{u2_1,u2_2,u2o}.  While 
conventional U(2) models provide an elegant theory of fermion masses, 
they do not generally lead to the large 2-3 down quark mixing of interest 
to us here.  In this section we present a minimal modification that yields
the desired result while preserving most of the desirable feature of
the original U(2) models.

We assume the flavor group $G_f=$~U(2)$\times$~U(1)$_F$, with the three
generations of matter fields again embedded in ${\bf 2}+{\bf 1}$
representations of the U(2) factor, and the symmetry breaking pattern
given in Eq.~(\ref{eq:brk}).  The additional U(1)$_F$ factor is assumed to
break at the first of the two symmetry-breaking scales in
Eq.~(\ref{eq:brk}), and is therefore associated with a small parameter
of size $\epsilon$. Conventional U(2) models involve singlet, doublet
and triplet flavon fields $A^{ab}$, $\phi^a$ and $S^{ab}$, that obtain
a pattern of vevs that are consistent with Eq.~(\ref{eq:brk}). In our
model, we make the unconventional choices $\epsilon\sim\lambda^2$,
$\epsilon'\sim \lambda^6$, and
\begin{equation}
\frac{\langle A \rangle}{M_f} \sim \left(\begin{array}{cc} 0 &
  \lambda^6 \\ -\lambda^6 & 0 \end{array}\right), \,\,\,\,\,
  \frac{\langle \phi \rangle}{M_f} \sim \left(\begin{array}{c} 0 \\
  \lambda^2 \end{array}\right), \,\,\,\,\, \frac{\langle S
  \rangle}{M_f} \sim \left(\begin{array}{cc} \lambda^6 & \lambda^6 \\
  \lambda^6 & 0 \end{array}\right) \,\,\, .
\end{equation}
Unlike the usual U(2) model, the first generation fermions get their
Yukawa eigenvalues primarily from the $(1,1)$ entry of the symmetric
flavon, rather than the off diagonals. Note that the $\lambda^6$
entries in the symmetric flavon are completely consistent with
Eq.~(\ref{eq:brk}); in the effective theory below the first symmetry
breaking scale, these entries correspond to fields with differing
charges under the intermediate U(1) symmetry acquiring order
$\epsilon'$ vevs.  All the entries are consistent with the dynamical
assumption that a given flavon will either get a vev of the same size
as a symmetry-breaking scale, or no vev at all.  In addition, we
introduce the following additional flavons with nontrivial U(1)$_F$
charge, indicated by a subscript:
\begin{equation}
\frac{\langle \phi_+ \rangle}{M_f} \sim \left(\begin{array}{c} 0 \\
\lambda^2 \end{array}\right),\,\,\,\,\, \frac{\langle \sigma_-
\rangle}{M_f} \sim \lambda^2 \,\,\,\,\, \frac{\langle \sigma_+
\rangle}{M_f} \sim \lambda^2 \,\,\, .
\label{eq:triv}
\end{equation}
Complete U(2) representation of the matter fields can be assigned
different U(1)$_F$ charges.  We assign a U(1)$_F$ charge of $-1$ to the
right-handed up-quark superfields of the first two generations and
$+1$ to the right-handed down quark superfield of the third
generation\footnote{We follow the usual convention that all matter fields
are embedded into left-handed chiral superfields.  Our charge assignments
apply to these fields.}.  We then obtain the Yukawa textures
\begin{equation}
Y_u \sim \left( \begin{array}{ccc} \lambda^8 & \lambda^8 & 0 \\
\lambda^8 & \lambda^4 & \lambda^2 \\ 0 & \lambda^2 & 1
\end{array}\right), \,\,\,\,\,\,\,\,\, Y_d \sim \left(
\begin{array}{ccc} \lambda^6 & 2.5 \lambda^6 & 0 \\ 0.5 \lambda^6 &
0.5 \lambda^4 & \lambda^4 \\ 0 & \lambda^2 & \lambda^2
\end{array}\right) \,\,\, ,
\label{eq:nonabyuk}
\end{equation}
and the soft scalar masses
\begin{equation}
m^2_Q \sim  m^2 \left( \begin{array}{ccc}
c_0 +\lambda^{12} & \lambda^{10} & \lambda^8 \\ \lambda^{10} & c_0 +\lambda^4 &
\lambda^2 \\ \lambda^8 & \lambda^2 & c_3 \end{array}\right), \,\,\,\,
m^2_U \sim m^2_D \sim \left( \begin{array}{ccc}
c_0 +\lambda^{12} & \lambda^{10} & \lambda^{10} \\ 
\lambda^{10} & c_0 +\lambda^4 & \lambda^2 \\ \lambda^{10} & \lambda^2 & c_3 
\end{array}\right)\,\, .
\end{equation}
Here, since the first two generation squarks are degenerate, we have
explicitly shown the ${\cal O}(1)$ coefficients in the diagonal entries (the
$c_0, c_3$'s are different for $Q,U, D$).  Also, while we generally
show only the order in $\lambda$ of a given Yukawa entry, we have
displayed some of the order one coefficients in $Y_d$.  This is to
illustrate that with a mild, ${\cal O}(\lambda)$ fine tuning between
$\langle A \rangle$ and $\langle S \rangle$ we can obtain the Cabibbo angle
from the down quark Yukawa matrix while sufficiently suppressing its 21 entry, 
thus avoiding the flavor changing problems discussed in Section II.

Finally it is worth pointing out that the similarity between the down quark
and charged lepton mass hierarchies suggests that under the flavor group
$L \sim D$ and $E \sim Q$ or $L \sim Q$ and $E \sim D$, where $\sim$ 
indicates identical flavor charge assignments. Detailed differences in the 
mass spectrum can be accommodated using the freedom to adjust order one 
coefficients. The first (second) choice implies large LH (RH) 2-3 
mixing in the charged lepton sector, independent of whether the
theory has any grand unified embedding.  The bi-large neutrino mixing
that is favored by the current data must therefore partly originate from
the neutral lepton sector.  While we do not explicitly investigate this
issue here, it is worth pointing out that Aranda, Carone and Meade~\cite{u2_2}
have shown that bi-large neutrino mixing can arise entirely from the neutral
lepton sector in non-Abelian models with $2$+$1$ representation structure, as
a consequence of the structure of the right-handed neutrino mass matrix and
the nonlinearity of the seesaw mechanism.  Non-Abelian models of the type we discuss 
here can in principle be generalized along these lines.  We refer the interested
reader to Ref.~\cite{u2_2}.

To conclude this section, we point out that the models we have constructed
can be tested in future collider experiments assuming that these models are relevant
for explaining the $B \rightarrow \phi K_s$ anomaly.  Both models involve significant
right-handed down squark mixing, which implies that the gluino mass must be
relatively light, in the $150-200$~GeV mass range, based on the numerical
results in Figure~3 of Ref.~\cite{murayama}.   Using the off-diagonal
squark masses predicted in each of our models, we may obtain lower bound estimates
on the mass of first two generation squarks, from $D$-$\bar D$ and $K$-$\bar K$ mixing
constraints.  These are shown in the Figure below, assuming a common mass for all squarks of
the first two generations.  While there is uncertainty in the order one coefficients in
such predictions, we can conclude qualitatively that the discovery of very light squarks
of the first two generation would disfavor the models discussed in this section.

\begin{figure}[ht]
\epsfxsize 3.3 in \epsfbox{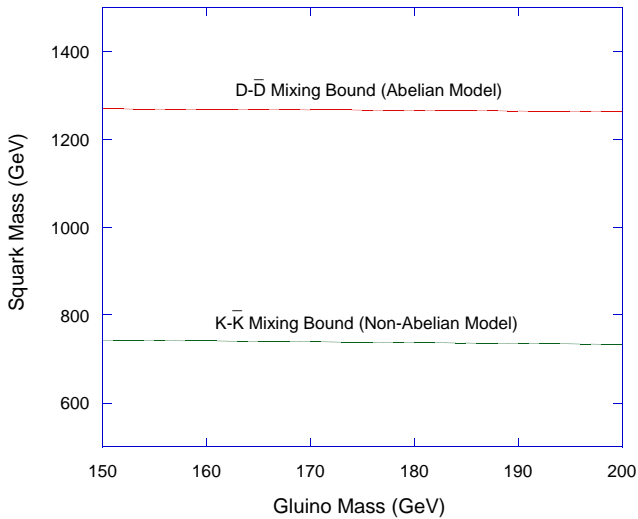} \caption{Typical bounds 
on the squark masses of the light two generations, assuming that 
unspecified order one coefficients are exactly 1, and the gluino 
is light enough to produce a significant effect in $B \rightarrow 
\phi K_s$.} \label{fig:new}
\end{figure}

\section{Conclusions}

We have considered the detailed flavor structure of supersymmetric
theories that can give large contributions to $b \rightarrow s$
transitions.  We have focused on the possibility that such theories
may explain the discrepancy between the value of $\sin 2\beta$
measured in $B\rightarrow \phi K_S$ and $B\rightarrow J/\psi \, K_S$
decays. With relatively few assumptions, we have isolated minimal,
preferred textures for the Yukawa couplings and the soft
supersymmetry-breaking masses.  In the case where the Cabibbo angle
originates in the down quark Yukawa matrix, we argued that the need
for squark degeneracy among the first two generations suggests an
underlying non-Abelian flavor symmetry with a ${\bf 2}\oplus {\bf 1}$
representation structure.  However, in a wide class of these models
with symmetric ${\bf 3}$ and antisymmetric ${\bf 1}$ dimensional
representations, the required, large 2-3 mixing is correlated with
Cabibbo-like 1-2 and 1-3 right-handed mixing angles. This leads
generically to unwanted flavor changing effects, given the requirement
that some elements of the superparticle spectrum must be light to
contribute non-negligibly to the processes of interest.  We argue that
non-Abelian models of this type may provide a viable solution to the
$\sin 2\beta $ anomaly providing that a mild fine tuning of parameters
is allowed.  On the other hand, if the Cabibbo angle originates in the
up quark Yukawa matrix, the same non-Abelian ansatz leads to a value
for the up quark mass that is too large.  Barring more complicated
non-Abelian constructions, the desired phenomenology seems in this
case to be realized more naturally in Abelian models that rely on
alignment rather than degeneracy to suppress strangeness-changing
neutral currents.  

We have presented explicit models that reproduce
many of the features of the idealized textures that we have discussed.
We showed how the minimal U(2) model, which normally does not have any
large mixing angles, may be modified with the help of an additional
U(1) factor to yield textures of the desired form.  In addition, we
presented a new type of alignment model with the desired properties
that is based on a single U(1) factor and a nontrivial extra-dimensional 
topography for the matter content.  If the $B\rightarrow \phi K_S$ anomaly 
turns out to be real, or if large deviations from SM predictions are seen in 
other $b \rightarrow s$ transitions, these models represent relatively 
minimal realizations of the desired flavor structure of the MSSM and 
provide a framework for investigating the correlation between a variety 
of flavor changing process in both quark and lepton sectors.




\begin{acknowledgments}
K.~A.~is supported by the Leon Madansky Fellowship and by
NSF Grant P420-D36-2041-0540.
C.D.C. thanks the NSF for support under Grant Nos.\ PHY-9900657,
PHY-0140012 and PHY-0243768.
K.~A.~thanks 
P.~Ko and M.~Piai
for discussions.
\end{acknowledgments}


\end{document}